# Robust magnetic anisotropy of a monolayer of hexacoordinate Fe(II) complexes assembled on Cu(111)

Massine Kelai,[†] Benjamin Cahier,[",○] Mihail Atanasov,[○,$] Franck Neese,[○] Yongfeng Tong,[†] Luqiong Zhang,[‡,"] Amandine Bellec,[†] Olga Iasco,["] Eric Rivière,["] Régis Guillot,["] Cyril Chacon,[†] Yann Girard,[†] Jérôme Lagoute,[†] Sylvie Rousset,[†] Vincent Repain,[†,*] Edwige Otero,[§] Marie-Anne Arrio,[‡,§] Philippe Sainctavit,[‡,§] Anne-Laure Barra,[⚹] Marie-Laure Boillot,["] and Talal Mallah[",*]

[†]Université de Paris, Laboratoire Matériaux et Phénomènes Quantiques, CNRS, F-75013, Paris, France
["]Institut de Chimie Moléculaire et des Matériaux d'Orsay, Université Paris-Saclay, CNRS, UMR 8182, 91405 Orsay Cedex, France
[○]Max-Planck-Institut für Kohlenforschung, Kaiser-Wilhelm-Platz 1, 45470 Mülheim an der Ruhr, Germany
[$]Institute of General and Inorganic Chemistry, Bulgarian Academy of Sciences, Akad.Georgi Bontchev street, Bl.11, 1113 Sofia, Bulgaria
[‡]Institut de Minéralogie, de Physique des Matériaux et de Cosmochimie, CNRS UMR7590, Sorbonne Université, MNHN, 75252 Paris Cedex 5, France
[§]Synchrotron SOLEIL, L'Orme des Merisiers, Saint-Aubin, 91192 Gif sur Yvette, France
[⚹]Laboratoire National des Champs Magnétiques Intenses, UPR CNRS 3228, Univ. Grenoble Alpes, 25, avenue des Martyrs, B.P. 166, 38042 Grenoble Cedex 9, France. E-mail: anne-laure.barra@lncmi.cnrs.fr

**Abstract**

The measurement of the magnetic anisotropy of [Fe{(3,5-$(CH_3)_2$Pz)$_3$BH}$_2$], where Pz = pyrazole, in its high spin $S$ =2 state by X-ray Magnetic Circular Dichroism (XMCD) spectroscopy when assembled as an organized monolayer on Cu(111) shows the presence of a hard axis of magnetization (positive axial zero-field splitting parameter $D$). Combining magnetization and multifrequency Electron Paramagnetic Resonance spectroscopy on a reference compound ([Fe{(3-(Ph)Pz)$_3$BH}$_2$]) of the same family and ab initio wave function based theoretical calculations, we demonstrate that the magnetic anisotropy of the assembled molecules is not affected when they are present at the substrate/vacuum interface. Comparing our results with those of a reported complex having almost identical FeN$_6$ coordination sphere but an easy axis of magnetization (corresponding to a negative $D$ value), we show that the nature of the magnetic anisotropy (easy/hard axis) is governed by the torsion angle ($\Psi$) defined by the relative orientation of the pyrazole five-membered rings to the pseudo three-fold axis of the molecules. The rigidity of the (Pz)$_3$BH tridentate ligands, where the three pyrazole moieties are held by the BH group, allows only very slight changes in the torsion angle even when the molecules are in a dissymmetric environment such as an interface. This is the origin of the robust magnetic anisotropy of this family of compounds.

**Introduction**

The study of how the magnetic properties of molecular magnets are modified when supported on a substrate is of great importance for their practical integration in solid state devices. Most of the molecular magnets studied in their bulk form are extremely sensitive to small deformations and it took some years to establish, by means of X-ray absorption spectroscopy (XAS), the quantum magnetization reversal for a monolayer of specially designed $Fe_4$ molecules grafted on an Au(111) surface.[1, 2] Regarding the change of magnetic anisotropy of such molecules when adsorbed on a surface, little is known. It has been shown recently that for planar molecules such as Fe(II)-phtalocyanin or Fe(II)-porphyrin, not only the magnetic anisotropy but also the spin state can be strongly modified on the surface, depending on the adsorption geometry,[3, 4] local distortions, and intermolecular interactions.[5] The origin of this large change in the spin electronic structure of this type of molecules is the relatively flexible and unsaturated coordination sphere of Fe(II). They possess, however, a large thermodynamic stability that allows keeping their integrity when sublimated under vacuum and assembled on metallic substrates. For non-planar molecular magnets, with a metallic ion in the most common hexacoordinate coordination sphere (distorted octahedral geometry), specially designed complexes have also shown a substrate-induced modification of their magnetic anisotropy because they were chemically grafted on oxide surfaces.[6] To ensure some robustness of magnetic anisotropy, a possible approach consists in using organic ligands that impose a relative rigidity on the metal ion coordination sphere so that major structural deformations are precluded when the molecules are present at the substrate/vacuum interface.

However, in order to investigate the magnetic anisotropy of a sub-monolayer of complexes, they must (i) keep their integrity when sublimed under vacuum and (ii) be able to assemble in an organized and not random (sub)monolayer on the substrate.

We have recently shown that the Fe(II) containing complex [Fe{(3,5-$(CH_3)_2$Pz$)_3$BH$\}_2$], where Pz = pyrazole (noted **1** in the following) that undergoes a complete thermal spin crossover in its bulk form,[7, 8] possesses all these requirements.[9] We demonstrated that a monolayer of **1** on Au and Cu presents only a partial thermal spin crossover (SCO) leaving molecules in the high spin (HS, $S$ = 2) state within the assembly,[9, 10] whose magnetic anisotropy can be investigated at low temperature. We, therefore, study the low temperature magnetic properties of a sub-monolayer (0.6 ± 0.15 ML) of **1** deposited on Cu(111) using X-ray Circular Magnetic Dichroism (XMCD). Because **1** cannot be in the HS state at low temperature in its bulk form (it undergoes a complete thermal SCO)[8] and because magnetic anisotropy is investigated at low temperatures, one cannot compare the magnetic properties of **1** in the two forms (bulk and as a monolayer) and cannot, therefore, draw conclusions on the effect of the substrate on its magnetic anisotropy. To circumvent this issue, we study the magnetic anisotropy of a reference complex [Fe{(3-(Ph)Pz$)_3$BH$\}_2$] (noted **2**), where the two methyl groups were replaced by an H atom and a phenyl group. Molecules of **2** remain high spin in the whole temperature range. In order to ensure that **2** is a reasonable benchmark of **1**, we carried out wave-function based *ab initio* theoretical calculations demonstrating that **1** and **2** have almost the same axial zero-field splitting (ZFS) parameters, therefore the same axial magnetic anisotropy. To ensure that theoretical calculations reproduce well the experimental data of **2**, we used complementary low temperature magnetization and multifrequency High Field Electron Paramagnetic Resonance (HF-EPR) spectroscopy to accurately determine the spin Hamiltonian parameters that describe its magnetic anisotropy.

We, first, present the description of the crystallographic structures of **1** and **2**, the experimental determination of the ZFS spin Hamiltonian axial (*D*) and rhombic (*E*) parameters for **2** in its bulk form, on one hand, and that of *D* of **1** assembled on Cu(111) (noted **1**/Cu), on the other hand. Then, we analyze the results of theoretical calculations justifying that **2** can be used as a reference for the magnetic anisotropy of **1** in its bulk form. Finally, we demonstrate that mainly *one* structural parameter governs the magnetic anisotropy of this family of complexes, a parameter that can barely be impacted by the molecules' environment due to the rigidity of the tridentate (Pz)$_3$BH capping ligands explaining the robustness of the magnetic anisotropy of **1** when present at the substrate/vacuum interface.

## Results and Discussion

### Synthesis and crystal structure of compounds 1 and 2

Compound **1** was prepared as already reported.[8] Its crystal structure in the $S = 2$ HS state was first reported at $T = 289$ K,[11] and then recently by some of us at $T = 298$ K.[8] The coordination sphere around Fe(II) (Figure 1a) is a slightly distorted octahedron. The Fe atom sits on an inversion center. Two Fe-N bond distances are almost identical (2.185 and 2.190 Å) and one is slightly different (2.160 Å). This is also the case for the $\widehat{NFeN}$ angles (for N belonging to the same tridentate ligand) equal to 86.22°, 86.3° and 87.5° (average value 86.7°). The $\widehat{BFeN}$ angles have an average value of 52.4° and differ by less than 0.5° (less than 1%). This angle value is only slightly lower than that of a regular octahedron (54.7°). The complex can then be structurally described as an elongated octahedron along a pseudo three-fold symmetry axis defined by the BFeB direction with a symmetry very close to D$_{3d}$.

Compound **2** was prepared using a procedure modified compared to that reported (see Methods).[12] The crystal structure of **2** (Figure 1b, Figure S1 and Table S1) is similar to that of **1**, but with some differences. The Fe atom sits on an inversion center. The Fe-N bond distances are larger than for **1** and have a larger deviation to their average value (2.210, 2.227 and 2.275 Å). The $\widehat{NFeN}$ angles have an average value very close to 90° (89.8°) larger than for **1**. The average value of the $\widehat{BFeN}$ angles (54.6°) is almost identical to that of a regular octahedron with a deviation less 0.2° (see Table S2).

Other relevant structural parameters are the torsion BFeNN angles ($\psi$) between the pseudo three-fold symmetry axis and the five-membered pyrazol rings of the tridentate ligand. For **1** and **2**, these angles hardly deviate from zero (from 1.65° to 5.69°, see Table S2) because they are imposed by the tridentate (Pz)$_3$BH capping ligands, they are almost the same for the two complexes.

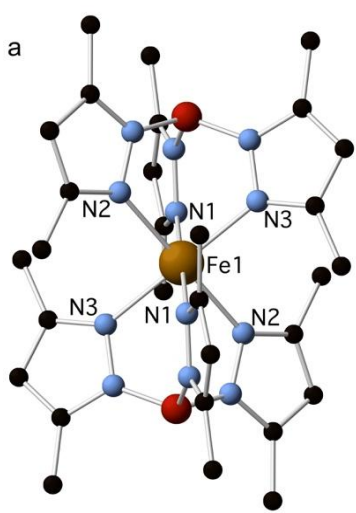

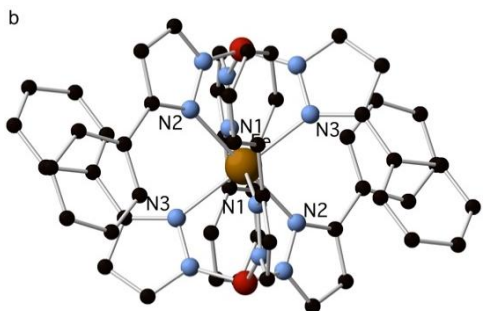

Figure 1. View of the molecular structures of **1** (a) and **2** (b), hydrogen atoms were omitted for clarity, N(blue), C (black), B (red), Fe (brown)

**Characterization of 1 assembled on the Cu(111) substrate (1/Cu)**

Davesne *et al.* first reported that **1** can be sublimed to form a ca. 600 nm film.[7] Then, some of us reported the formation of a sub-monolayer of **1** on gold[9] and later on other metallic substrates among them Cu(111) (see Methods) where an anomalous HS to LS light induced crossover was discovered at low temperature.[13] **1** was sublimed on Cu(111) substrate as already reported.[13] Figure 2 shows a typical small scale STM image of such a sample (see Methods). The molecules organize in large islands of a perfectly ordered two-dimensional molecular lattice with lattice parameters $\vec{A}$ = 8.7 ± 0.3 Å, $(\vec{B})$ = 10.6 ± 0.4 Å, $(\vec{A},\vec{B})$ = 89.5 ± 3°. A careful examination of the STM images shows that the molecules have two distinct absorption geometries that look mirror symmetric with respect to the $(\vec{A},(Oz))$ plane, and alternating along the $\vec{A}$ direction.

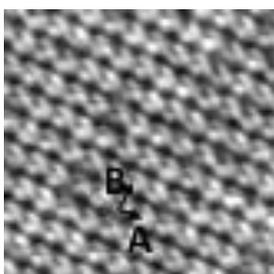

Figure 2. 10x10 nm² STM image of the periodic network of **1** on Cu(111) at 4.7 K. $V_t$ = 0.3 V, $I_t$ = 20 pA.

**Magnetic properties of 2**

The magnetic properties of **2** were measured using a SQUID magnetometer in the 2-250 K temperature range for $\chi T$ and the 0 - 5 T magnetic field for magnetization (see Methods). The thermal variation of the $\chi T$ product between 250 and 2 K shows the expected behavior for a $S$ = 2 state (Figure 3a). Upon cooling, $\chi T$ is constant (3.28 cm³ mol⁻¹K) down to 60 K corresponding to $S$ = 2 with a $g$-factor equal to 2.09, then it decreases to 0.9 cm³ mol⁻¹K at $T$ = 2 K. The magnetization (*M*) vs. the applied magnetic field (*B*) was measured at $T$ = 2, 4 and 6 K (Figure S2). The *M* = f(*B/T*) are not superimposable, they do not fall on one master curve as for a Brillouin function, which is the signature of non-negligible ZFS (Figure 3b). The simultaneous fit of the $\chi T$ = f(*T*) and the *M* = f(*B/T*) curves using the PHI package,[14] and considering the following spin Hamiltonian:

$$H = g\beta \vec{B}.\vec{S} + D(\hat{S}_z^2 - \frac{S(S+1)}{3}) + E(\hat{S}_x^2 - \hat{S}_y^2) \quad (1)$$

where $\vec{S}$ is the spin operator ($S_i$ (*i = x, y,z*) its components), $\vec{B}$ the magnetic field, ß the Bohr Magneton and *D* and *E* the axial and rhombic ZFS parameters respectively, gives the following parameters $g_2$ = 2.10, $D_2$ =

+11.4 cm$^{-1}$ and $|E_2|$ = 1.2 cm$^{-1}$ ($|E_2|/D_2$ = 0.1). No reasonable fit could be obtained with a negative $D$ value. These values of the ZFS parameters lead to the low-lying energy spectrum depicted in the inset of Figure 3a (see SI for the energy of the $M_S$ sub-levels for a $S$ = 2 state), with the $M_S$ = 0 lying lower which corresponds to a situation with a hard axis of magnetization ($D > 0$).

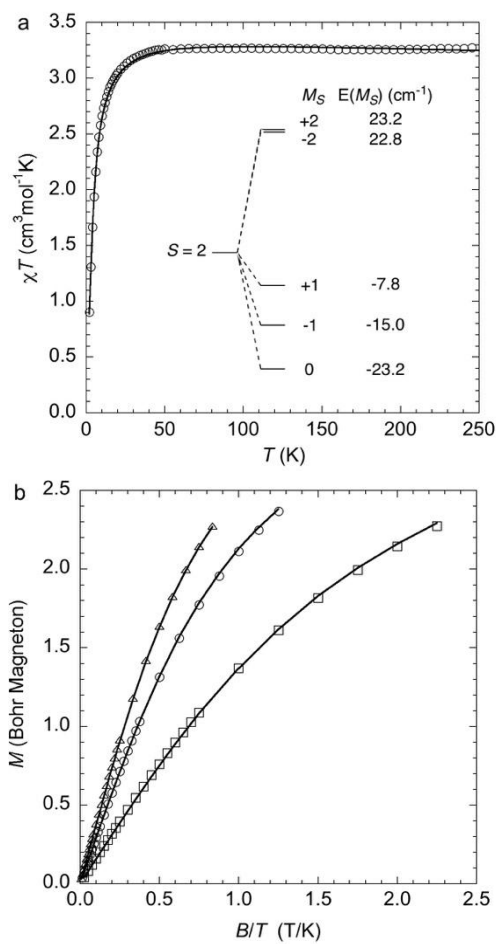

Figure 3. (a) Thermal variation of $\chi T$, (inset) energy spectrum of the $M_S$ sub-levels due to ZFS of $S$ = 2 and (b) magnetization ($M$) vs. $B/T$ at $T$ = 2 (□), 4(○) and 6 (△) K, the continuous lines correspond to the best fit (see text for parameters)

**HF-HFEPR on 2**

The multifrequency HF-HFEPR measurements (see Methods) revealed several signals for most of them, mostly for ν > 200 GHz (Figure 4a). The analysis of the resulting data suggests that molecules with different anisotropies (resulting probably from slight geometrical changes) are present. Indeed, as can be seen in the 662.4 GHz spectra (Figure 4a), the strong and well-resolved low field signals (between 4 and 6 T) correspond to forbidden transitions (off-axis turning points) whereas the signals appearing at higher field which correspond to allowed transitions are much broader, extending for some of them over 1 T. This points out to the existence of a distribution of anisotropies leading to broad signals, with most probably a non-gaussian distribution resulting in the well resolved low field signals. Considering a spin Hamiltonian where the isotropic $g$ of equation (1) is replaced by the **g** matrix, the analysis of the spectra recorded in the 190-662.4 GHz frequency range allowed obtaining a set of parameters defining the magnetic anisotropy for the molecules responsible of the signals observed close to 5.5 T at 662.4 GHz: $D_{2EPR}$ = 12.1 cm$^1$, $E_{2EPR}$ = 0.61 cm$^{-1}$ ($E_{2EPR}/D_{2EPR}$ = 0.05) and $g_x$ = 2.23, $g_y$ = 2.14, $g_z$ = 2.2. These values allow to reproduce several observed transitions (Figure 4b) and are in good agreement with those issued from the magnetic data analysis (susceptibility and magnetization). The $g_z$ value is the most uncertain one as no well-defined transition could be associated to the z orientation. Unfortunately, it is not possible to obtain a full set of parameters for the magnetic anisotropy of the molecules leading, for instance, to the 4.2 T signal at 662.4 GHz. An axial magnetic anisotropy $D'_{EPR}$ = 14.1 cm$^{-1}$ is found considering a slightly larger rhombicity ($E'_{EPR}$=1.14 cm$^{-1}$ i.e. $E'_{EPR}/D'_{EPR}$ = 0.08) with the same $g$ values; these values are thus indicative of the anisotropy. The allowed transitions associated to this minority species are compatible with the experimental linewidths of the allowed transitions and agree with a non-gaussian distribution of anisotropies. These results thus indicate that $D$ is positive and close to 12-14 cm$^{-1}$, and that the rhombicity is low. It also shows (as usually found by EPR, an extremely sensitive technique) that there is a weak distribution of the ZFS parameters reflecting a slight distribution in the structure of the molecules within the bulk compound.

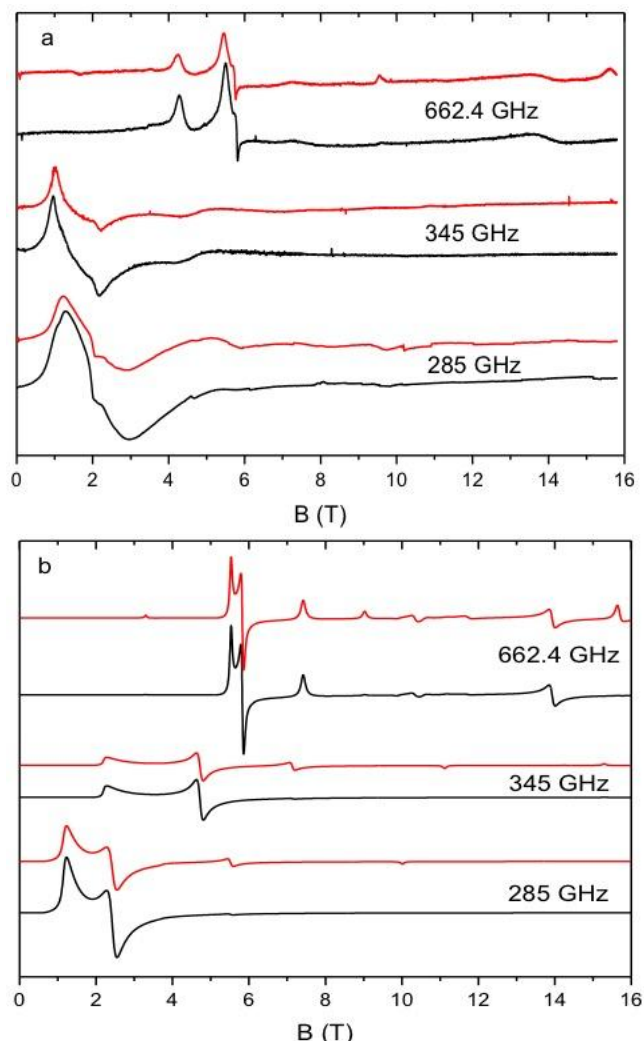

Figure 4. Experimental (a) and calculated (b) EPR powder spectra at three different frequencies and at 5 K (black) and 15 K (red).

The anisotropy of **2** being determined by magnetometry and EPR studies, we focus below on **1** assembled on Cu(111).

**XAS and XMCD on 1/Cu(111)**

The molecular layer of **1** was slowly-cooled down to 2 K and investigated by X-ray Absorption Spectroscopy (XAS) and X-ray Magnetic Circular Dichroism (XMCD). The XAS data show the presence of a mixed HS/LS state with a HS proportion of 75.0 ± 0.7%.[13] The sample was then irradiated by X-rays that allows, due to the SOXIESST (Soft X-rays Induced Excited Spin State Trapping) effect to perform a LS to HS transformation with time.[15, 16] After typically two hours of illumination, the sample reaches a steady state with 90% fraction of the HS state as demonstrated by the XAS spectrum displayed in Figure 5a and Figure S3. By using left ($\sigma^-$) and right ($\sigma^+$) circularly polarized light and high magnetic field (6.0 T), we can measure the corresponding XMCD spectrum (Figure 5b) that gives a direct information on the magnetic properties of the molecular layer.

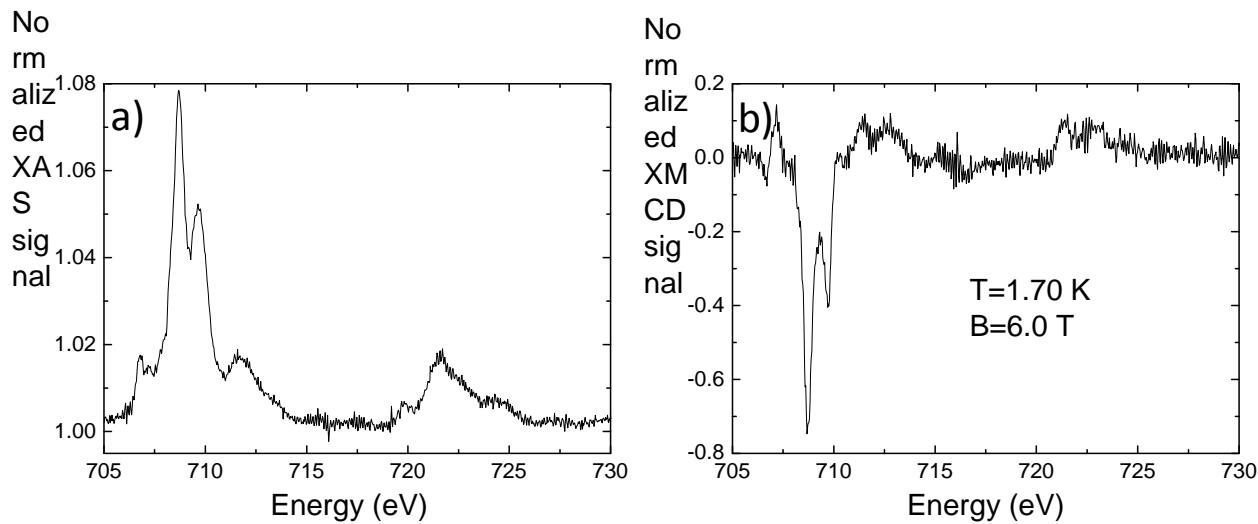

Figure 5. a) Averaged XAS spectrum (*T* = 1.70 K, *B* = ±6.0 T, σ± polarities) recorded after two hours of X-ray illumination, showing a 90% HS character. A linear background has been removed and the spectrum has been normalized by its background value. The jump at the edge can therefore be read directly in percentage, b) XMCD spectrum (*T* = 1.70 K, *B* = 6.0 T). The spectrum has been normalized by the corresponding XAS jump at the *L₃* edge. The signal can therefore be read in percentage of the maximum XAS signal.

First, we can extract by sum rules analysis, the ratio of the orbital moment over the spin moment.[17, 18] This ratio, independent of the number of holes, added to the electron gyromagnetic number $g_0$, gives the so called effective gyromagnetic number that takes into account the remaining orbital moment in the spin Hamiltonian formalism. By measuring the sum rules at three different temperature, we find $g_{eff}$= 2.2 ± 0.2. We can also measure the maximum intensity of this spectrum, at 708.7 eV, as function of the applied magnetic field. We therefore obtain curves directly proportional to the magnetization curves. Figure 6 displays for three different temperatures the averaged measurements over two full magnetic cycles (from -6 T to +6 T and from +6 T to -6 T), where the error bars correspond to the standard deviations of the measurements. In order to better emphasize the difference with a Brillouin paramagnetic behavior, we have represented this signal as function of the ratio B/T, for which the three sets of data should collapse into the same curve in the absence of magnetic anisotropy in the molecule. To obtain quantitative values on the anisotropy axis, we have fitted those data by using the PHI software, in a spin Hamiltonian framework,[14] with $g_{1/Cu}$ fixed to 2.2, the value found by the sum rule. The fitting parameters are $\theta$, the angle between the magnetic field and the principal anisotropy axis of the complex, $D_{1/Cu}$ is the axial ZFS parameter and finally a normalization factor as the XMCD signal is only proportional to the magnetization. The full lines (Figure 6) show the best fit over the three temperatures data, giving $\theta$ = 50 ± 10° and $D_{1/Cu}$ = 7.6 ± 1 cm⁻¹.

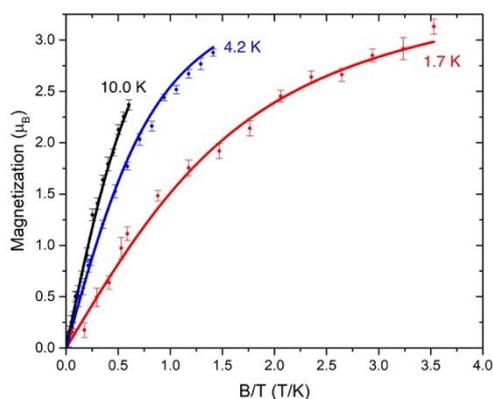

Figure 6. Data points represent XMCD signal at 708.7 eV of 0.15 ML of **1** on Cu(111) as function of $B/T$ where $B$ is the magnetic field in Tesla (from 0 to 6 T) and $T$ the temperature in Kelvin (red dots at 1.7 K, blue dots at 4.2 K and black dots at 10.0 K). The full lines represent the best fit curves for $g_{1/Cu}$ fixed to 2.2, $D_{1/Cu}$ = 7.6 ± 1 cm$^{-1}$ and θ = 50 ± 10°. A normalization factor has been applied to the XMCD intensity in order to fit the magnetization curves.

As stated in the introduction, we performed theoretical calculations to compare the magnitude and the nature of the magnetic anisotropy of the two complexes and to check whether **2** can be considered as a reasonable reference for **1**.

**Theoretical calculations**

Wave function based theoretical calculations were performed including dynamic correlation using the ORCA 4.2 package.[19] The crystallographic structural data were used for the calculations, after optimizing the positions of hydrogen atoms by DFT. The calculations give, for the two compounds, almost the same value for the axial ZFS parameter ($D$) and a slightly different value for the rhombic one ($E$): $D_{1calc}$ = 8.8 cm$^{-1}$, $|E_{1calc}|$ = 0.62 cm$^{-1}$ ($|E_{1calc}|/D_{1calc}$ = 0.07) and $D_{2calc}$ = 10.0 cm$^{-1}$, $|E_{2calc}|$ = 1.1 cm$^{-1}$ ($|E_{2calc}|/D_{2calc}$ = 0.11). The principal values of the **g** matrix were found to be almost identical for the two complexes ($g_{1x}$ = 2.00, $g_{2y}$ = 2.10, $g_{3z}$ = 2.12 and $g_{2x}$ = 1.99, $g_{2y}$ = 2.12, $g_{2z}$ = 2.17), close to the average values found for **2** from magnetization data. The calculations reproduce quite well the experimental data of **2**. They are suitable to account for the value of the axial ZFS parameter of **1** in its high spin state and can, therefore, be compared to the experimental value for the molecules assembled on Cu(111). The results are summarized in Table 1.

Table 1. Axial ZFS parameter values calculated[c] and measured for **1**/Cu and **2** from different techniques

| $D$ (cm$^{-1}$) | **2** | **1**/Cu |
|---|---|---|
| Magnetometry/XMCD | +11.4[a] | 6.6 - 8.6[b] |
| EPR | 12 - 14 | - |
| Calculations | 10.0 | 8.9[c] |

[a] from Squid measurements, [b] from field dependent XMCD measurements, [c] calculated for **1**

The main result is that the magnetic anisotropy of molecules of **1** is almost not altered ($D_{1/Cu}$ and $D_{1calc}$ are equal to 7.6 and 8.9 cm$^{-1}$ respectively) leading to the *preliminary* conclusion that the molecules do not sustain large structural deformation even when adsorbed on a metallic substrate. However, in order to get a deeper insight in the effect of structural deformation on the magnetic anisotropy of **2**, one needs to determine which structural parameters affect most magnetic anisotropy, which can be done by analyzing the results of the ab initio calculations.

The first result concerns the orientation of the ZFS **D** tensor axes that provides information on the orientation of the magnetic axes with respect to the molecular frame. The calculations show that **D** tensor axes are colinear for the two compounds (Figure 7, Figure S4). The principal axis (z) is found along the pseudo three-fold axis direction B-Fe-B of the two complexes and the y and x axes deviate by less than 2°. The location of the hard magnetization axis (z, blue in Figure 7) of **2** allows to propose an orientation of the molecules on Cu(111) with the pseudo three-fold axis making an angle of 50 ± 10° with the normal to the plane (40 ± 10° with the substrate). θ, in this study, can be compared to a reported one obtained from a structural study of the same molecule deposited on Au(111).[9] In this latter case, an angle close to 70°(20° with the substrate) was found but both the two-dimensional lattice and the STM images on Au(111) were different from the ones measured on Cu(111) (Figure. 2). Only a grazing incidence X-ray diffraction study on this system could confirm in the future this particular orientation of the molecules with respect to the substrate.

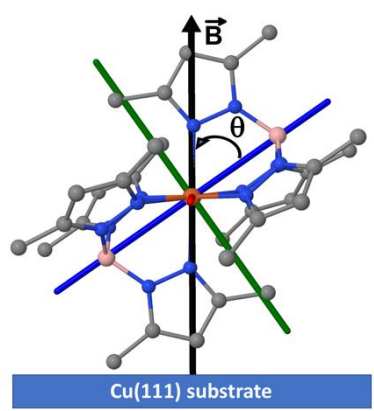

Figure 7. **D** tensor axes for **1** (x (red), y (green) and z (blue)), the magnetic $\vec{B}$ is shown parallel to the normal to the substrate making an angle θ of 55° with the z axis (blue) as found from the fit of the magnetic data of **1**/Cu(111). The drawing gives, therefore, the orientation of the molecule on Cu(111).

The second aspect concerns the rationalization of the positive value of *D*, which will allow us to determine the structural parameters that most impact the magnetic anisotropy. The slight differences in the structural parameters of **1** and **2** (Fe-N bond distances and $\widehat{NFeN}$ angles, see Table S2) barely affect the axial ZFS parameter *D*, and the slight change on the rhombic parameter is not significant due to the uncertainties on its calculation. The orientation of the principle anisotropy axis and its (hard) nature are the same for the two compounds. This can be understood by examining the electronic structures of the two compounds that are very similar given by the state (Figure S5) and the molecular orbital (MO) energy diagrams (Figure 8) obtained from ab initio calculations.

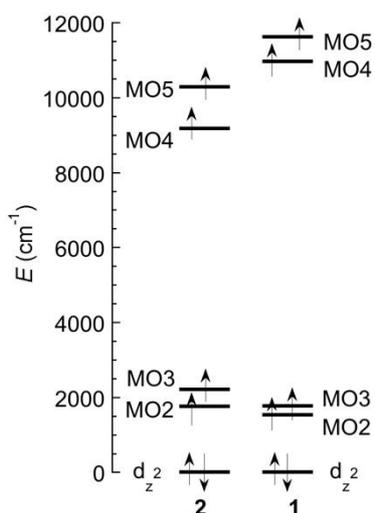

Figure 8. Energy diagram of the one electron molecular orbitals for **1** and **2** obtained from ab initio ligand field theory (AILFT) calculations by ORCA.

For **1** and **2**, the orbital energy diagrams follow the expected scheme for a trigonal symmetry ($D_{3d}$ for example) where the d orbitals transform as the irreducible representation $A_{1g}$ ($z^2$), $e_g$ (xz, yz) and $e_g$ (xy, $x^2-y^2$). We find that the lowest MO is a pure $d_{z^2}$ orbital, then a set of two nearly degenerate MOs (MO2 and MO3) very close in energy at around 2000 cm$^{-1}$ and then another set of two orbitals (MO4 and MO5) with close energies at around 11000 cm$^{-1}$. The composition of the different MOs is given in Table S3. The splitting within the two sets (MO2, MO3) and (MO4, MO5) reflects the deviation from trigonal symmetry. The $d_{z^2}$ orbital is almost non-bonding. Indeed, it has zero overlap with the σ like nitrogen atoms orbitals that point almost exactly in its nodal cone (Figure 9a), the $\widehat{BFeN}$ angles being very close to 54.7°. In addition, the π-like orbitals of the N atoms are almost orthogonal to $d_{z^2}$ because the torsion angles (ψ, see Table S2) are very close to 0° also leading to an almost zero overlap (strict orthogonality is obtained for ψ equal to 0° as depicted in Figure 9b).

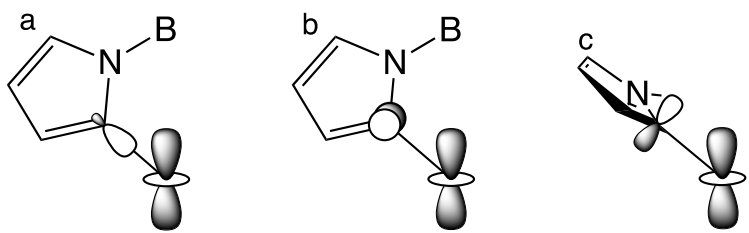

Figure 9. Scheme of the potential interaction between the $d_{z^2}$ orbital on one hand and nitrogen atoms orbitals coordinated to Fe: (a) with the σ axial orbital for ψ = 0°, (b) with the π-like orbitals for ψ = 0° and (c) for = 90°, showing in this last case the presence of non-zero overlap rendering the $d_{z^2}$ orbital antibonding.

Considering the electronic structure of the two complexes and particularly comparing the composition of the ground ($Q_0$) and excited quintet states (Table S4 and Figure S5), it is possible to rationalize the sign of *D* for the two complexes. The positive values of the *D* parameters are due to the mixing via spin orbit coupling of the ground quintet state ($Q_0$) with mainly the two first excited states ($Q_1$ and $Q_2$). The ground state is expressed by a single slater determinant ($|xy,yz,xz,x^2-y^2|$ in the hole formalism) and the composition of the excited ones is given in Table S4. One notices that these two excitations ($Q_0$ to $Q_1$ and $Q_0$ to $Q_2$) lead to a change of ±1 of the value of the orbital momentum $m_l$ which corresponds to a positive contribution to *D* (a more detailed analysis is given in the SI).[20, 21] In summary, the fact that the $d_{z^2}$ orbital has the lowest energy is responsible of the positive *D* values for **1** and **2**, and small differences in the bond distances and angles cannot much affect its energy relative to MO2 and MO3 and cannot, therefore, much affect the magnitude of *D*. Only if the relative energies of $d_{z^2}$ on one hand and MO2 and MO3 on the other hand is reversed, a negative *D* value could be obtained. Such change occurs only if ψ changes from close to zero to a large value. Indeed, in such a case, the first order orbital angular momentum $m_l =|1|$ in the nearly degenerate ground state ($^5$E) and the fact that the excitation between the ground and the first excited states involves no change in the value of the orbital momentum lead to a negative contribution to *D*.[20, 21] Such change in the energy of the $d_{z^2}$ orbital requires an overlap with the π orbital of the nitrogen atoms that may occur if the value of ψ changes. One can already suggest, at this stage, that a large change of ψ in the family of complexes to which **1** and **2** belong is unexpected because the tridentate (pz)$_3$BH ligand imposes a rigid geometry around Fe precluding a deviation by more than a few degrees from zero. We examined the Cambridge Crystallographic Data Centre and ref.[22] and found that for hexacoordinate complexes with any derivative of the (pz)$_3$HB tridentate ligand and different metal ions (Fe$^{II}$, Co$^{II}$ and Ni$^{II}$), the ψ values are never larger than 9°. This supports a rather high degree of rigidity against deformation leading to the conclusion that the axial anisotropy of the family of [Fe$^{II}$(pz)$_3$BH)$_2$] related complexes should not be easy altered by their environment.

However, in order to check the proposed model we examined the molecular structure of a Fe(II) complex ([Fe(ptz)$_6$]$^{2+}$, ptz is propyl tetrazole, see Figure S6 for a view of the structure) similar to **1** and **2** that was shown to have a negative *D* value form EPR (X-band and high field, *D* = -14.8 cm$^{-1}$).[23] If one considers only the *first* coordination sphere FeN$_6$ the three complexes (Table S2 and ref.[24] for the structure of [Fe(ptz)$_6$]$^{2+}$), one finds that they are very close. The negative *D* value for [Fe(ptz)$_6$]$^{2+}$ cannot, therefore, be justified only on this basis. The examination of the torsion angles of [Fe(ptz)$_6$]$^{2+}$ (defined here as the angle between the three-fold axis of the molecule and the plane of the tetrazole five-membered ligand) reveals that they are completely different, they are equal to 63.3° for [Fe(ptz)$_6$]$^{2+}$ (Figure S6), while between 1.7° and 5.7° for **1** and **2**. In order to prove that the change in magnetic anisotropy from hard to easy axis is mainly governed by this torsion angle, we performed ab initio calculations on [Fe(ptz)$_6$]$^{2+}$ that give a negative *D* value and therefore an easy axis of magnetization for the complex (Figure S7). The MO energy diagram (Figure S8 and Table S5) show that the orbital with a majority contribution from $d_{z^2}$ does not have the lowest energy anymore as expected from the onset of an overlap with the nitrogen atoms π orbitals due to the change of the torsion angle as depicted in Figure 10. This is responsible of the negative to *D* value in this complex.

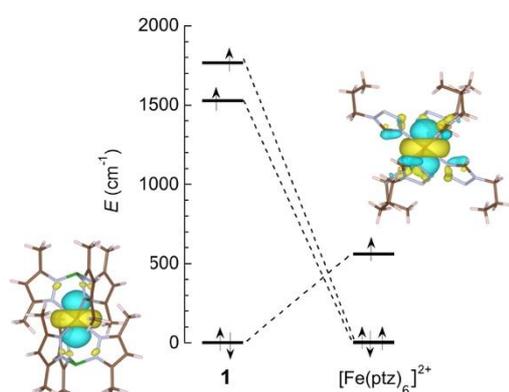

Figure 10. Relative energies of the three low-lying MOs for **1** and [Fe(ptz)$_6$]$^{2+}$, with the sketch of the MOs having a majority contribution from $d_{z^2}$ (see SI, Figure S9).

**Concluding remarks**

We have measured the magnetic anisotropy of the high spin (*S* = 2) state of [Fe{(3,5-(CH$_3$)$_2$Pz)$_3$BH}$_2$] molecules assembled on Cu(111). These data were compared to a reference complex ([Fe{(3-(HPh)Pz)$_3$BH}$_2$] whose anisotropy was determined by magnetization and HF-HFEPR spectroscopy studies. They are consistent with a magnetic anisotropy of the complexes that barely changes when in contact with the metallic substrate. The examination of the structure of a reported complex with a very similar structure but with a negative $D$ value leads to the conclusion that the main structural parameter that affects the axial magnetic anisotropy of these systems are not the bond distances and angles in the FeN$_6$ *first* coordination sphere of Fe but the degree of rotation of the pyrazole ring about the Fe-N bond (defined by the torsion angle ($\psi$) between the (pseudo) threefold axis of the molecules and the five membered rings). When $\psi$ is close to zero, $D$ is found to be positive and rationalized by wave function based theoretical calculations. While, when this angle is much larger (63° as for [Fe(ptz)$_6$]$^{2+}$), $D$ is found to be negative.[23, 24] For **1** (and **2**), because the five membered pyrazole rings of the tridentate ligands (3,5-(CH$_3$)$_2$Pz)$_3$BH coordinated to Fe cannot rotate freely about the Fe-N bond and cannot deviate by more than few degrees from 0 when the complex is placed in a dissymmetric environment, the axial magnetic anisotropy cannot be *highly* affected as demonstrated here when the molecules are organized at the metal/vacuum interface. However, as shown by the examination of the structure of a series of complexes belonging to this family, the torsion angle can vary within a very small range (from 0° to 9°), which may slightly affect the magnetic anisotropy. A distribution of the values of the $\psi$ (there are six within a complex) among the molecules within a sample may not be detectable by X-ray diffraction on single crystals, but may affect the EPR spectra that is sensitive to tiny variation of the ZFS parameters. This is, to our mind, the origin of the distribution observed in the EPR study. Finally, this study suggests a route of preparing Fe(II)-containing hexacoordinate complexes with an easy axis of magnetization that can be assembled on metallic surfaces without an alteration of their anisotropy providing the use of organic ligands that impose a $\psi$ angle as close as possible to 90°.

**Methods**

*Preparation of the monolayer of 1/Cu(111)*

The monolayer of **1** on Cu(111) has been realized and measured in ultra-high vacuum chambers (base pressure from low $10^{-11}$ to low $10^{-10}$ mbar). The Cu single crystal is cleaned by cycles of sputtering (Ar$^+$ pressure between 5.10$^{-6}$ mbar, 900 eV) followed by a standard annealing at 850 K. Its cleanness has been checked by Auger Electron Spectroscopy and STM prior to deposition. The powder of **1** is thermally sublimated from a home-made quartz Knudsen cell at around 85 °C. Depending on the samples, the deposition is done with the substrate at 4.7 K followed by an annealing at room temperature or a direct deposition on the substrate at room temperature. We have checked by STM that the two deposition techniques give the overall same morphology of the molecular layer. The Cu(111) sample has been prepared and measured by STM on a separate UHV chamber, before being transported to the DEIMOS UHV setup in a home-made vacuum suitcase with a pressure below $10^{-9}$ mbar during typically three hours. A XAS spectrum at the Cu $L_3$ edge at 930 eV shows no trace of contamination. The estimation of the coverage of **1** on Cu(111) is done by comparing the normalized jump at the edge with a reference Cu(111) sample covered with 0.4 ± 0.1 ML that has been determined by STM images.[13] The STM measurement has been performed on a low temperature (4.7 K) Omicron STM setup.

*Preparation and characterization of* [Fe{(3-(Ph)Pz)$_3$BH}$_2$] ((3-(Ph)Pz)$_3$BH is hydrotris(3-phenylpyrazol-1-yl)borate) (**2**)

All manipulations are conducted under anaerobic atmosphere. A 15 ml methanolic solution of K(HB(3-Ph-Pz)$_3$) (300 mg, 0.624 mmol in ) is added dropwise to a 15 ml solution of Fe(BF$_4$)$_2$•6(H$_2$O) (105.3 mg, 0.312mmol). A white precipitate immediately appears. It was filtered, dried under vacuum and then recrystallized from a THF solution. Single-crystals of X-ray quality of Fe(HB(3-Ph-Pz)$_3$)$_2$ were isolated by slow evaporation of a THF solution.

Elemental analysis Calcd for $C_{54}H_{44}N_{12}B_2Fe$: C, 69.11; H, 4.73; N, 17.91; Found: C, 68.72; H, 4.88; N, 17.69. IR ($\nu$/cm$^{-1}$): 3124 (w), 3060 (m), 3028 (w), 2500 (w), 2470 (m, B-H), 1519 (m), 1498 (s, CN), 1464 (s), 1446 (m), 1369 (s), 1341 (m), 1303 (w), 1276 (m), 1252 (w), 1206, 1189 (vs), 1126 (m), 1115 (m), 1081 (w), 1072 (m), 1057, 1049 (s), 1029 (m), 998 (m), 980 (w), 952 (w), 920 (w), 907 (w).

*X-ray Absorption Spectroscopy*

The XAS spectroscopy is performed with circularly polarized X-ray with an energy resolution of 150 meV. The total incoming photon flux is typically $10^8$ photons.s$^{-1}$.mm$^{-2}$. The sample is placed with the X-ray incident direction along the surface normal which is also the direction of the applied magnetic field. A total electron yield mode (drain current measurement) is used to detect the Fe(II) $L_2$ and $L_3$ edges. XMCD spectra are done by an average difference of circularly right and circularly left polarization and positive and negative magnetic fields.

*Magnetic measurements*

Magnetization measurements were performed on a Quantum Design MPMS5 SQUID magnetometer operating the 2-300 K temperature and 0-5 T magnetic field ranges. The powder obtained from ground crystals of **2** was blocked in parafilm to avoid any torqueing effects. Data was corrected from parafilm contribution and diamagnetism was estimated from Pascal constants.

*HF-HFEPR*

The HF-HF EPR measurements were carried out on a microcrystalline powder sample of **2** pressed into a pellet to reduce torqueing effects. A multi-frequency spectrometer operating in a double-pass configuration was used.[25] Frequencies were provided by 95 and 115 GHz Gunn oscillators (Radiometer Physics GmbH) and a 110 GHz frequency source (Virginia Diodes Inc.) associated to multipliers up to the sixth harmonic. The detection is performed with a hot electron InSb bolometer (QMC Instruments). The exciting light is propagated with a corrugated waveguide inside the cryostat and with the help of an optical table outside. The main magnetic field is supplied by a 16 T superconducting magnet equipped with a VTI (Cryogenic). The measurements were done on powdered samples pressed into pellets in order to limit torqueing effects. Calculated spectra were obtained in two steps: a fitting of the identified resonance positions,[26] to obtain the parameters driving the spin Hamiltonian (eq. 1) and a calculation of the spectra with the SIM program.[27] Both programs were developed by H. Weihe (Univ. of Copenhagen).

*Theoretical calculations*

All calculations have been done with the Orca 4.2 package.[19] Optimization of hydrogen position was done with DFT using B3LYP functional and the def2-TZVP basis set. For computational speed-up the phenyl groups of (2) were replaced by hydrogen atoms.

The *D* and *E* parameters were evaluated following the procedure developed in ref.[28] A State Average CASSCF (Complete Active Space Self Consistent Field) was performed; then, the dynamical correlation is added by NEVPT2 method in its strongly contracted scheme, without frozen core.[29-31] Finally, the Spin-Orbit (SO) coupling was accounted for by quasi-degenerate perturbation theory with the SOMF Hamiltonian.[32] The Complete Active Space (CAS) is composed of the five mainly-3d orbitals of the Fe ions and the 6 associated electrons, *i.e.* CAS(6,5). The averaging of the molecular orbitals CASSCF optimization (MO) was done over all the 5 quintets and 45 triplet spin states generated by the CAS(6,5). The SO coupling was considered between all these states, the spin-free energy (diagonal elements of the SO matrix) being evaluated at the NEVPT2 level. In CASSCF, NEVPT2 SO calculations, the relativistic Ahlrichs basis set were used (DKH-def2-TZVP for Fe atom, DKH-def2-TZVP(-f) for B, C and N atoms and DKH-def2-SVP for H atoms).[33]

**Acknowledgement**

This project has received funding from the European Union's Horizon 2020 research and innovation program under grant agreement No [766726]. This work was supported by the Centre National de la Recherche Scientifique (CNRS, France) and Université Paris-Saclay. TM thanks the Institut Universitaire de France (IUF) for financial support.

**Author contributions**

M.K., L.Z., Y.T., V.R., A.B., E.O., M.-A. A. and Ph. S. performed and analyzed the STM, XAS and XMCD. J.L., Y.G., S.R., C.C contributed to the discussion on STM results. M.K. and V. R. performed the fit of XMCD results. O. I. and M. L-B. prepared the compounds and characterized them. E.R. carried out the magnetic studies on bulk compounds. B.C., M.A and F.N. performed and analyzed the theoretical calculations. A.-L. Barra performed and analyzed the EPR. data. V.R. and T.M. wrote the paper.

**Conflicts of interest**
There are no conflicts of interest to declare

Supporting Information Available from

**Graphical abstract**

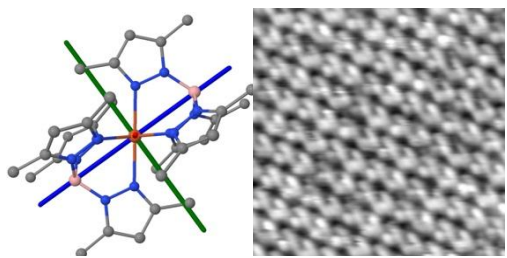

The tridentate ligand tris pyrazolyl borate imposes a rigid organic scaffold around Fe(II) preventing the twist of the five membered pyrazole group, therefore ensuring a robust magnetic anisotropy when the molecules assembled as monolayers suffer from the dissymmetric environment of the substrate/vacuum interface.